# APPRAISE: a governance framework for innovation with AI systems

Diptish Dey[1] and Debarati Bhaumik[2]
Amsterdam University of Applied Sciences, The Netherlands
{d.dey2, d.bhaumik}@hva.nl

Abstract: As artificial intelligence (AI) systems increasingly impact society, the EU Artificial Intelligence Act (AIA) is the first serious legislative attempt to contain the harmful effects of AI systems. This paper proposes a governance framework for AI innovation. The framework bridges the gap between strategic variables and responsible value creation, recommending audit as an enforcement mechanism. Strategic variables include, among others, organization size, exploration versus exploitation -, and build versus buy dilemmas. The proposed framework is based on primary and secondary research; the latter describes four pressures that organizations innovating with AI experience. Primary research includes an experimental setup, using which 34 organizations in the Netherlands are surveyed, followed up by 2 validation interviews. The survey measures the extent to which organizations coordinate technical elements of AI systems to ultimately comply with the AIA. The validation interviews generated additional in-depth insights and provided root causes. The moderating effect of the strategic variables is tested and found to be statistically significant for variables such as organization size. Relevant insights from primary and secondary research are eventually combined to propose the APPRAISE framework.

Keywords: AI governance, responsible AI, AI act, audit

## 1 Introduction

The symbiosis between humans and AI is up for a new trial, fairness and discrimination, driven by ethical zeal (Bringas Colmenarejo et al., 2022) and legislative will (Mökander et al., 2022). The large-scale proliferation of AI systems within a relatively short span of time has been accompanied by their undesirable impact on society, which has become common knowledge (Angwin et al., 2016). For example, it takes the forms of price discrimination in online retail based upon geographic location (Mikians et al., 2012), and in unsought societal impact of algorithmic pricing in tourism and hospitality (van der Rest, 2022). Deploying classification models these systems, due to their increasing complexity, are often uninterpretable by humans. The mounting struggle to fully comprehend the rationale behind decisions made by these systems, fuels the need to examine and scientifically explain such rationale (explainability) (Miller, 2019) and subsequently evaluate these systems from fundamental rights and ethics (Ashok et al., 2022) perspectives. The need for a more structured, comprehensive, and actionable approach (Kazim & Koshiyama, 2021) is screaming for attention, catering to which should involve perspectives beyond legislation and ethics only.

The calls for responsible AI combining governance (Stahl, 2022; Merhi, 2023), mechanisms, and means of participation have been mounting, driven by discrimination/biasness concerns of individuals, ethical considerations of psychologists and social activists, corporate social responsibility (CSR) objectives of corporations and the responsibilities of legislators to protect fundamental rights of citizens (Rai, 2022). These stakeholders, and probably more, may be operating independently, but at an organizational level the problem is interconnected in their concerns. This paper characterizes these concerns within four pressures, technology, value creation, regulatory, and normative, in sections 2, 3, 4, and 5 respectively. That organizations perform balancing acts when subjected to similar pressures has been observed earlier (Winecoff & Watkins, 2022). Through an empirical study presented in section 6, and two validation interviews presented in section 7, this paper illustrates how organizations experience these pressures. Subsequently, section 8 presents APPRAISE, a novel governance framework for AI innovation that assists organizations embracing AI for product/service innovation to audit themselves. Section 9 presents high level conclusions and directions for future research.

---

[1] https://orcid.org/0000-0003-3913-2185
[2] https://orcid.org/0000-0002-5457-6481

## 2 Technology pressure

Withstanding the many trials and tribulations since its humble beginnings a century ago (Floridi, 2020), AI and its applications are increasingly entrenching deeper and wider into our social fabric. In its earliest applications to solve business problems, the approach was mathematical and rule-driven (Holloway, 1983). The focus was on creating innovative algorithms or improving existing ones. The turn of the century witnessed the emergence of Big Data (Boyd & Crawford, 2012), and subsequently data-driven AI, which in hindsight can be viewed as a natural consequence of the convergence of technological developments as enablers of AI (Brynjolfsson & McAfee, 2017), and the wide acceptance of these technologies (Sohn & Kwon, 2020).

The relationship between an organization's ability to innovate and its financial viability has been investigated by Bai & Tian (2020), and Eisdorfer & Hsu (2011). In the backdrop of organizations leveraging AI to discover and implement new business models (Lee et al., 2019), the consequence of not embracing AI is potentially fatal for organizations. In business model innovation, technology has traditionally been viewed as an external antecedent (Foss & Saebi, 2017). Limiting AI to an external antecedent merely, grossly underestimates the comprehensive nature of its impact: AI strengthens the way other technologies perform, amplifying their added value. In creating value, AI transforms businesses and overall economic systems (Soni et al., 2020; Mazali, 2018). Organizations are confronted with a pressure to embrace AI.

This pressure reveals itself in a sense of urgency to embrace AI. How much of that urgency translates to true urgency (Kotter, 2008) as opposed to complacency and false urgency needs to be studied within organizations realizing innovations (Mitcheltree, 2023). Additionally, the speed at which innovation is realized is essential to its viability due to costs and market relevance. When innovating with AI, the need for speed often necessitates interdepartmental exchanges (cross-border and/or cross-cultural), offshoring and collaboration, which need to be managed (Sarin & Mahajan, 2001). Change management becomes challenging. The "considerable length of time" that Kotter (1995) proposes is unrealistic, since the speed of innovations in AI is unprecedented (Tang et al., 2020). Furthermore, the cross-cultural, cross-border nature of AI innovations makes organizational alignment (Kotter, 2008) challenging and time-consuming. It is this combination of urgency and challenges in change management that aggravates the technology pressure.

## 3 Value creation pressure

Business model innovation, as discussed earlier, involves strategic reviews in organizations aimed at developing strategic plans to capture value from AI innovations (Wamba, 2022). Empirical research and insights into how such strategic plans eventually create strategic value are limited in literature (Borges et al., 2021). Broadly speaking, AI-driven innovation adds value to organizations in three ways (Bahoo et al., 2023): through product/ service innovation (Bhardwaj, 2021), process innovation (Ghahramani et al., 2020), and/or improved decision making (Araujo et al., 2020). In realizing value within these three contexts, challenges that organizations encounter at different phases of their innovation roadmaps (Phaal et al., 2001) or strategic plans differ based on their attributes.

The continuum of organizations includes small startups to large multinationals operating within markets that vary in their degree of maturity. To what extent organization size moderates successful AI innovation, is a subject of research. For larger organizations, the exploitation/exploration dilemma (March, 1991) implies choosing between "ambidexterity" (Benner & Tushman, 2015) and "punctuated equilibrium" (Burgelman, 2002). Despite the preference in literature for ambidexterity (Mishra & Pani, 2021), it is worthwhile to note that in organizational units, in which exploration is prevalent, the price of over-rigid processes is lack of innovation. The discussion is less relevant for startups, where the scope of AI is narrower, as compared to larger organizations, in which diverse applications of AI could be exploited and/or explored.

In innovating with AI, an organization's dilemma to build or buy, often results in parts of the application development process outsourced (Gerbert et al., 2018), regardless of organization size (Mariani & Fosso Wamba, 2020). Networked innovation outsourcing (Guan & Wang, 2023) frequently results in organizations offshoring their development activities. How the level of outsourcing and offshoring influences an organization's eventual success in embracing AI is a subject of research. Communication across these multiple interfaces is challenging (Piorkowski et al., 2021). In coordinating value creation and creating accountability in this diverse context, organizations may draw parallels from software engineering (Nguyen-Duc & Cruzes, 2013). One major difference between AI innovation and traditional software engineering is in the approach towards compliance: whereas in the latter, compliance has been driven by standards (Disterer, 2013), models/metrics (Kan, 2003), and individual/organizational/ cultural aspects (Mubarkoot et al., 2023), the former has the additional complexity of (upcoming) regulation influencing how organizations embrace AI innovations to create value.

## 4 Regulatory pressure

Key technology-related legislative developments in the EU over the last decade include the widely acknowledged GDPR (Voigt and Von dem Bussche, 2017) and the proposed AIA (Commissie, E., 2021). The GDPR aims at protecting the privacy of individuals, against a backdrop of by and large benevolent and bulging data processing, often facilitated by AI algorithms. The GDPR assists data controllers, processing personal data, in achieving compliance. Similarly, the AIA presents a conformity regime, in which accountability is directed to participants in the whole supply chain of algorithms, from their conception to deployment and during their life cycle.

### 4.1 Initial effects of GDPR

Parallels have been drawn between GDPR and ISO 27001:2013; conforming to the latter, automatically enables compliance on many GDPR provisions (Lopes, 2019). However, unlike the GDPR, the ISO is very action-minded and auditable with its prescribed controls and control objectives. Organizations, legal experts, and courts experience GDPR as difficult to interpret, apply and lacking in enforcement mechanisms (Saqr, 2022).

The impact of GDPR on online interaction between consumers, web technology providers, and websites, has been investigated by Peukert (2022). The research draws interesting and relevant conclusions. First, the number of requested third party domains that respond with cookies when an EU citizen browses a website hosted in the EU, has dropped about 10% during the GDPR introduction in April 2018. Second, market concentration of web technology providers increased post GDPR, with the bigger service providers increasing market share at the cost of the smaller ones. This implies that the decline in the usage of third-party cookies affected the smaller technology providers more than the larger ones. The work of Peukert is best summarized as: an undesired heterogenous effect, a relatively small decline in third-party cookies, and an impact with possibly a short-term memory.

## 4.2   The proposed EU AI Act

The European Commission proposed the AIA, which presents a conformity regime of regulating AI systems (Schuett, 2023). The AIA plays the role of a gatekeeper through ex-ante conformity assessments and associated penalties, which is a commonly accepted role within regulation. Additionally, it describes the role of enforcement through its prescription of post-market monitoring systems. In doing so, the AIA provides clarity not only on the proposed legislation, but also on the initial key steps needed to regulate it through a regulatory body (the notified body).  Using a tiered risk-based approach, the AIA defines three tiers, "low- or no-risk", "high-risk" and "unacceptable risk". High-risk systems are subject to ex-ante conformity assessments and post-market monitoring systems driven by competence in specific AI technologies. The AIA stipulates that a provider of a stand-alone high-risk AI system can choose to conduct ex-ante conformity assessment internally if its AI system is fully compliant or use an external auditor if its AI system is either partially compliant or harmonized standards governing that AI system is non-existent.

The AIA addresses accountability in the supply chain and encompasses various standpoints, among others, prohibited practices, transparency obligations, governance, and compliance procedures across the supply chain of these systems (Bhaumik & Dey, 2023). It "*sets out the legal requirements for high-risk AI systems in relation to data and data governance, documentation and recording keeping, transparency and provision of information to users, human oversight, robustness, accuracy and security*" (Commissie, 2021, p. 13). Articles 10 to 15 of Chapter 2 Title III of the AIA elaborates on these legal requirements.

Efficacy of the proposed AIA and its economic cost have been questioned and criticized by some studies (Mueller, 2021), which present an unfavourable future environment in the EU, an environment that is decoupled from the global AI market. Nevertheless, the increased interest in AI ethics across the wide range of stakeholders point to more efficient and effective regulatory frameworks in the future (Bringas Colmenarejo et al., 2022). Furthermore, the focus will shift towards operationalizing the regulatory frameworks with a view to making them more actionable, not only by the regulator, but also at the organizations that are regulated.

# 5   Normative pressure

That technology leads to undesirable social effects is not new (Nye, 2006). Yet, the concern today is *maximi momenti* for three reasons. First, history of discrimination stands to repeat itself, unprecedented in its global scale, its sweeping impact due to its pace and breadth and, its newly found cohort, uncontrolled AI systems. Second, blindfolded by its bountiful benefits the problem appears to have been underestimated; nevertheless, the proposed legislations by the EU are highly welcome steps in the right direction. Third, holding a creator of a self-learning AI system accountable for all its future actions is infeasible based on the *Principle of Alternative Possibilities* (Frankfurt, 2018). As a result, anxious groups distancing themselves and unable to reap the benefits of these systems are increasingly common (Swant, 2019). That participation of all stakeholders is important for responsible innovation has been researched by Jasanoff (2016).

Without much ado, it is fair to state that the complexity of defining the scope of AI and therefore what constitutes as an AI profession is immense (Gasser & Schmitt, 2020). The influence of AI is felt across all sectors and contexts (Chui et al., 2022). On the one hand, this includes medical doctors and lawyers in traditional professions with high norms dosed through and advocated by their respective associations that oversee their professional standards. On the other hand, the amateurish, experimental nature of building machine learning models has resulted in its enormous popularity (Simonite, 2018), which from a professional standards perspective has no fundament to build upon. Driven among others by corporate social responsibility, organizations such as Google and Microsoft have formulated their own AI ethics and norms (Pichai, 2018; Microsoft, 2018) whereas a professional association such as ACM has proposed an aspirational code of conduct that could be used to charter their own professional norms (ACM, 2018). The demarcation between ethics and norms is often unclear in these standards and in the absence of an industry-wide set of norms, there is often too much room for AI professionals and organizations to manoeuvre. This often leads to employees protesting the lack of professional norms at work (Nieva, 2022).

Individual moral responsibility (IMR) plays a key role in ensuring that employees act ethically and morally (Poel et al., 2015). When individuals perceive themselves as morally responsible, they are more likely to engage in ethical behaviour (Yunianti et al., 2019 and take responsibility for any negative consequences that may arise out of their actions. This sense of accountability can motivate individuals to comply with rules and regulations, even when there may not be any external pressure to do so. Important mediators to IMR include freedom to act (Talbert, 2021), causality (Hwang et al., 2019), and culpability & moral sense (Rudy-Hiller, 2022). The absence of industry-wide professional norms, and therefore, a deficit in collective moral responsibility, implies that organizations need to demonstrate leadership in creating the pre-conditions that enable IMR in their AI innovation teams. With increasing complexity and growing popularity of self-learning systems, it becomes essential that users and providers of AI-systems inspire ethical conduct, for example through stimulating IMR (Royakkers et al., 2015), since, it has been observed that organizations may unknowingly increase unethical behavior as a byproduct of their pursuit of high performance (Mitchell et al., 2018).

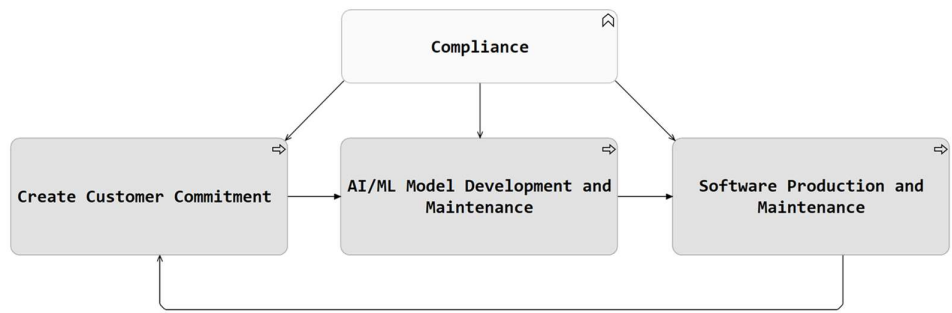

*Figure 1: Organizational paradigm*

## 6   Empirical research to measure adherence to the AIA

With an eye on proposing a governance framework, an empirical study was deemed necessary to analyse the extent to which organizations comply with the AIA and its compliance products. The latter was further restricted to understanding to what extent organizations coordinate technical elements of AI systems to ultimately comply with the AIA. Early results from the experiment, involving a smaller set of participants and limited analyses, have been presented by Walters et al. (2023). Of the three ways that AI systems add value to organizations (see section 3), the empirical research was limited to organizations embracing AI for product/ service innovations.

Figure 1 presents an organizational paradigm, in which functional requirements are defined within the Create Customer Commitment (CCC) process and passed onto the AI Model Development and Maintenance process, which in turn ensures the development, monitoring and update of AI models. These models are finally deployed within the Software Production and Maintenance process. The KPIs for these processes differ, e.g., the CCC process is driven by commercial KPIs, whereas, in model development, prediction accuracy as a KPI is common practice. The Compliance function serves the three processes ensuring that legal obligations are refined and met, processes are audited and eventually accountability is created.

## 6.1 Design of the experiment

The material scope of the empirical research was finalized by performing an analysis of the AIA and its implications on functional and technical requirements of AI systems. The scope was further restricted to functional requirements as opposed to technical ones. The distinction between functional and technical requirements is driven by a couple of reasons. Unlike in GDPR, where compliance can be achieved through focus on technical requirements (Lopes et al., 2019), such as security and privacy, AI-systems operate closer to the core business of an organization, directly affecting decisions within its primary processes. Additionally, in our opinion, technical requirements serve as qualifiers to the AIA compliance of a system, whereas functional requirements emanating from the CCC process could be implemented from a wide selection of algorithms, depending upon an organization's risk appetite. It is this variability that creates differentiation in the primary process between organizations. Nevertheless, it is worthwhile mentioning that the choice of functional requirements (and therefore choice of algorithms) may have consequences on technical requirements (Khatter & Kalia, 2013). In the drive to create value (and therefore financial gains), functional requirements are variables that are likely to be optimized, sometimes in ways that could be judged as unethical. The analysis of the AIA resulted in an overview of 12 elements that could be grouped into 8 topics (see table 1).

*Table 1: Elements from the AIA that impact functional requirements directly*

| Topics | Elements extracted directly from the AIA |
|---|---|
| Data | Training/validation and test data should be sufficiently relevant, representative, free of errors, and have complete and appropriate statistical properties. |
| | Assessment of availability, quantity, and suitability of the data. |
| | Examination of possible bias. |
| | Identification of possible data gaps or shortcomings and how to address these. |
| Technical documentation | Assessment of human oversight measures and technical measures to facilitate output interpretation. |
| | Validation and testing procedures used metrics to measure accuracy, robustness, cybersecurity, and compliance. |
| User communication | Instructions of use, including information around possible risks to rights and discrimination, and the level of accuracy should be. |
| Human oversight | A natural person must oversee the system and have the ability to take action that cannot be overwritten, and this person must have the necessary competence, training, and authority. |
| Accuracy | Systems should perform consistently and meet an appropriate level of accuracy in accordance with the state of the art. |
| Monitoring | Learning after placement must be accompanied by rules establishing that changes to the algorithm and its performance have been assessed and pre determined by the provider. |
| Risk management | Must be established, implemented, documented, and maintained. This is an iterative process requiring regular updating. |
| Quality management | A quality management system should be established for governance purposes. |

The topics in table 1, a result of the AIA analysis, were further grouped into five topics: data & model, technical documentation, user application, model monitoring and model risk. Such a grouping was driven by the need for easier and structured communication, when accumulating primary data through interviews with professionals. Furthermore, topics (and their elements) were overlapping, and an attempt was made to make the elements in table 1 mutually exclusive and collectively exhaustive. Besides, the elements as stated in AIA often have consequences to varying degrees on the technical requirements of the system. If it could be argued that a specific part of an element is more a technical requirement than a functional one, it was decided to exclude it from the current scope.

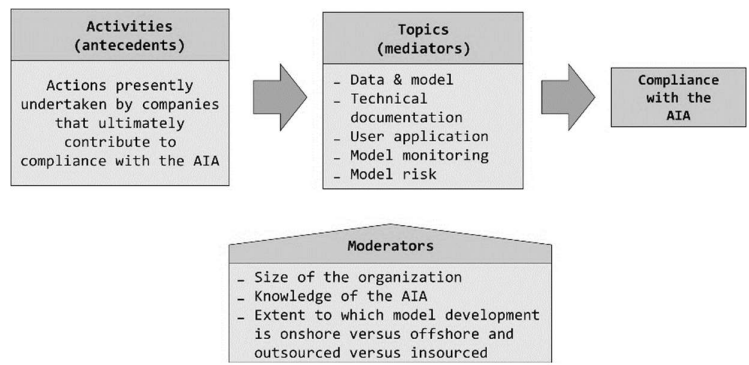

*Figure 2: Experimental design*

The experimental design is depicted in figure 2, where the five topics are depicted as mediators. The activities that organizations currently undertake within these five topics are antecedents that contribute to the level of compliance that these organizations ultimately demonstrate with the AIA.

Four moderators were selected: size, prior knowledge of the AIA and the extent to which organizations outsource or offshore their model development. Since the AIA proposes sandboxes for startups and because size plays an important role on how organizations create value (section 3), size was an important moderator. To understand how regulatory pressure (section 4) influences compliance, it was essential to consider 'prior knowledge of the AIA' as a moderator. Finally, the level of outsourcing and the level of offshoring were chosen as moderators because the AIA addresses accountability across the whole AI supply chain and because they play important roles in how organizations create value (section 3).

## 6.2 Methods

This section presents an overview of the methods and techniques used to create a survey, select participants to be interviewed, ensure desired outcomes of interviews, and analyse the outcomes.

### 6.2.1 The survey

The survey was structured in such a way that participants would first have to answer the moderator questions and then respond to the mediator questions. It was not possible to go back and change responses to questions. It was possible for the surveyor to make additional notes, such as listing questions in which the participant asked clarifications and the time spent on answering each of the five topics was also noted. The survey was developed and rolled out using Qualtrics between April 2023 and October 2023. Organizations in the Netherlands were targeted due to their geographical proximity.

The five topics within the defined scope were populated with closed questions as indicated in table 2. A total of 76 closed questions spread over these five topics were developed. A Likert scale was used to collect the responses. Depending upon the question and the possible answers to that question, the Likert scale varied between 2-points and 6-points. In many questions, respondents could also voluntarily leave their comments behind within an open text field. Additionally, the questions could be categorized as either "activities" or "perception". The activity questions (in total 49) correspond to antecedents in the experimental design of figure 2. These questions identify if an organization undertakes specific activities that count towards AIA compliance. The perception questions (in total 27) aim at identifying to what extent an organization supposes that it is compliant.

The moderators accounted for five questions, also measured using a Likert-scale. These questions are presented at the bottom of table 2. Given the limited number of questions per moderator, generating insights based solely upon these questions would have been unreliable. The interactions during the mediator questions added valuable insights into how accurate participants had been in their responses to the moderator questions. Two additional questions were included in the survey to understand the extent to which organizations feel the pressure to comply with the AIA.

### 6.2.2 Participant selection

The focus of the survey was the Netherlands with (social media) networking and interviewing potential candidates as the primary method of sourcing respondents. Checks were made to ensure that the organizations selected were proactively developing an AI-system as part of their product offering like the paradigm depicted in figure 1. The range included organizations at various stages in building and deploying AI-systems: from very recently (less than a year) launched systems with relatively simple algorithms to very complex algorithms deployed for more than five years. This would ensure that the organizations had hands-on experience with topics from the AIA. Conscious effort was made to include large multinationals and startups in the survey. During participant selection, it was difficult to select based upon the other moderators: knowledge of the AIA and the level of outsourcing/offshoring as it was difficult to measure these without diving into the survey questions. Nevertheless, only participants who were senior enough in their roles to be to certain extent held accountable for the AI systems they developed were selected as opposed to junior developers, whose focus lay on model/software development.

### 6.2.3 Interview protocols

Online interview was the preferred method of guiding respondents through the survey questions. The respondents had access to the online Qualtrics link, which they either filled-in on their own or filled-in while sharing their screen in a MS Teams environment. The latter accounted for more than 90% of the respondents. Respondents were informed verbally and in writing prior to taking the survey that their response to the survey was confidential, implying that their identity and that of their organization would never de disclosed. They had to fill in their name, their organization, and their email address as a pre-requisite to taking the survey questions. Those who chose to fill in the survey on their own without guidance were considered valid responses if they disclosed their identity and that of their organization, which would be verified later. Responses, where this verification was not possible, were considered invalid for the experiment. It was estimated that responding to the survey would take between 45 minutes to an hour. Filling-in the survey with guidance was strongly preferred and indicated to the participants. This was done with the purpose of reducing ambiguity when interpreting the questions and for the interviewers to get a detailed picture of the organization in relation to the experiment. A minimum of 3 interviewers were scheduled for each interview. With an eye on reliability, post interview, the interviewers would reflect upon the responses, especially with relation to the moderator questions.

*Table 2: List of survey questions*

| Topics | Questions |
| --- | --- |
| Data & model - Activity | When does your organization check the suitability (e.g., correctness) of the data for the purpose of the system? |
| | How often does your organization recollect the data when it is not suited for the purpose of the system? |
| | How often does your organization adjust data (e.g., append or delete features) when it is not representative of the purpose of the system? |
| | How often does your organization adjust data when it is found to contain errors? |
| | Has your organization noticed risks associated with using a dataset in the past 2 years? |
| | How often does your organization mitigate risks in a dataset? |
| | At which moment in the data gathering phase are actions taken to mitigate risks concerning the dataset? |
| | Which stakeholders do you involve in the process of risk mitigation? |
| | Does your organization train its employees on data and model bias? |
| | Which departments are trained on data and model bias? |
| | How often do you test model performance per demographic? |
| | How often are performance test results on demographics communicated with stakeholders? |
| | With which stakeholders are test results communicated? |
| | How often does it occur that a dataset misses data? |
| | How often does a data gap lead to an action, in which the existing dataset is adjusted, or new data is gathered to fix the data gap? |
| | With which stakeholders are data gaps communicated? |
| | How often does your organization reflect upon the acceptable level of model accuracy? |
| Data & model - Perception | My organization always ensures that data used for the purpose of the system is highly suitable. |
| | My organization always ensures that data used for the purpose of the system is representative and free of errors. |
| | My organization identifies and mitigates risks associated with a dataset. |
| | My organization involves stakeholders in the process of risk mitigation. |

|  | | |
|---|---|---|
| | | Professionals in my organization are aware that data can be biased. |
| | | Professionals in my organization are aware that bias can occur within the model. |
| | | Are these professionals, working on a project, aware of its customer demographics? |
| | | Model performance on different demographics is tested and communicated all across the system development value chain. |
| | | We are expected to follow a protocol if it comes out that the dataset misses data. |
| | | Within my organization data gaps are communicated to different stakeholders. |
| | | My organization has internally accepted levels of what a good accuracy is. |
| Technical documentation - Activity | | Are there dedicated professional(s) for writing technical documentation in your organization? |
| | | With which stakeholders is technical documentation shared? |
| | | Technical documentation of my organization is written for technical people only |
| | | There are guidelines within my organization to ensure completeness of technical documentation. |
| | | How often is technical documentation written? |
| | | At which stage in the development process is technical documentation written? |
| | | How often are compliance requirements in relation to technical documentation communicated with the development chain? |
| | | With which stakeholders are compliance requirements in relation to technical documentation communicated? |
| | | Is someone in your organization trained to determine the compliance requirements of the technical documentation? |
| | | Are there guidelines for writing the technical documentation within your organization? |
| | | How often are guidelines for technical documentation revised? |
| Technical documentation - Perception | | Technical documentation is very detailed within my organization. |
| | | The entire system development value chain is responsible for making technical documentation of the system. |
| | | Technical documentation is shared with the management of my organization. |
| | | In my organization multiple departments understand the technical documentation. |
| | | In my organization compliance requirements for technical documentation are communicated across the system development value chain. |
| | | My organization adheres to clear policies when writing technical documentation. |
| User application - Activity | | With whom does your organization communicate accepted risks of the system? |
| | | How often does your organization communicate accepted risks of the system with stakeholders? |
| | | How does your organization estimate a model's risk on rights and discrimination? |
| | | How often do you measure a model's risk on rights and discrimination? |
| | | How often does your organization test that the customer of the system understands what the model is predicting? |
| | | How often does your organization perform user (e.g., customers) tests? |
| User application - Perception | | My organization communicates accepted risks of the system with the user of the system. |
| | | My organization has metrics to estimate a model's risk on rights and discrimination. |
| | | My organization tests that the user of the system understands what the model is predicting. |
| Model monitoring - Activity | | How many professionals in your organization are responsible for monitoring the AI systems? |
| | | What type of qualifications do people who monitor the systems have? |
| | | My organization has a dedicated employee, who is appropriately qualified, for monitoring the systems. |
| | | How often is model monitoring performed? |
| | | How often are actions taken post monitoring? |
| | | Post monitoring, how often do you check if model requirements are still met? |
| | | What kind of actions do you take post monitoring, if the model does not meet requirements anymore? |
| Model monitoring - Perception | | Are model monitoring and model development the responsibility of the same department? |
| | | Models that learn after deployment are monitored to ensure that the model still meets requirements. |
| | | My organization has clear quality standards to which a system must adhere. |
| System risk management - Activity | | How many different employees are directly involved with managing risks related to the model/ system? |
| | | Which departments/professionals are involved in the risk management system? |
| | | How often is the risk management system assessed? |
| | | How often does your organization take a corrective action after a risk has been flagged? |
| | | How often do you test that a corrective action has mitigated the risk? |
| | | How often are flagged risks actually mitigated? |
| | | How often are risks of the system communicated across stakeholders? |
| | | With which stakeholders are risks of the system communicated? |

| System risk management - Perception | My organization has established a risk management system. |
| --- | --- |
| | My organization assesses risk management at regular intervals. |
| | My organization coordinates risk mitigation across the entire value chain of system development. |
| | My organization communicates risks of the system across all key stakeholders. |
| Moderator – Organization Size | How many employees does your organization have? |
| Moderator – Knowledge of the AIA | How would you rate your level of familiarity with the Act? |
| | Have you received any training or guidance on the Act? |
| Moderator – Insourcing/ Outsourcing | To what extent does your organization "insource" and/or "outsource" development of AI models? |
| Moderator – In-shoring/ Offshoring | To what extent does your organization develop AI models "inshore" and/or "offshore"? |
| AIA pressure | Do you believe that the AIA will have a significant impact on your organization's use of AI? |
| | How much of a priority does your organization currently place on complying with the AIA? |

## 6.3 Analyses

The activity questions in table 2 were used to calculate the actual compliance score, whereas the perception questions were used to calculate the perceived compliance score. Each question and each of the five topics had equal weightage in these two scores. The responses within the Likert-scales were standardized in the range [0,1]. A yes/no, was converted to a 0/1 scale. In questions where more than a single response was possible, points were distributed per response and the total points were standardized in the range [0,1] by dividing it with the maximum possible points. This way the actual compliance scores and the perceived compliance scores could be calculated within each dimension of the survey for each participant. Dividing the perceived compliance score by the actual compliance score provided the exaggeration ratio. The exaggeration ratio reflects the extent to which a participant was overstating its actual compliance level.

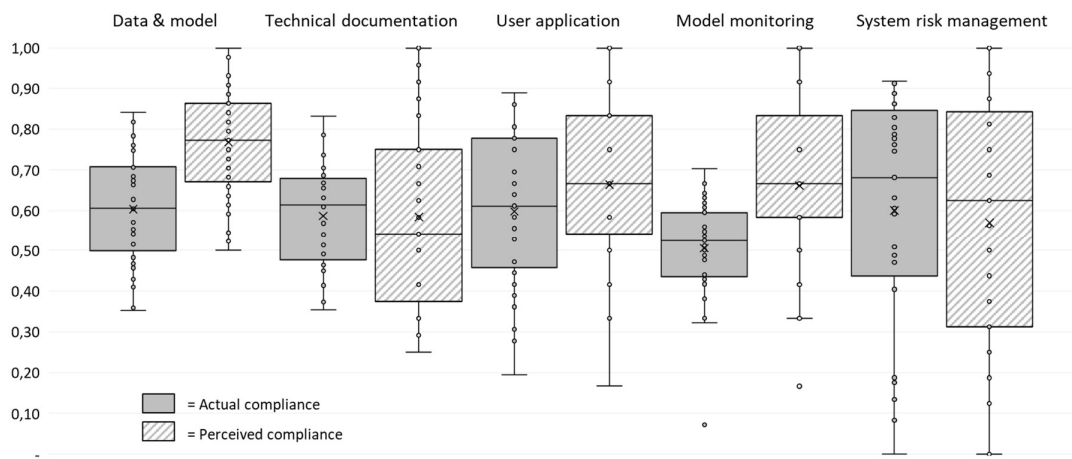

*Figure 3: Overview of actual and perceived compliances of the participants*

### 6.3.1 Compliance differs per topic (mediator)

In the 6 months that the survey was open, 42 organizations responded, of which 34 organizations successfully completed the survey. All the 8 organizations that are excluded, decided to take the survey on their own without assistance from the interviewers; the second half of their survey responses was typically empty. Of the 34 selected organizations, 33 were guided by the interviewers. Their responses are summarized in the boxplots depicted in figure 3, which shows the mean, median and variance of the 34 responses within the five topics and split by the actual and perceived compliance questions. Except for model monitoring (0.51), the means of actual compliance of the four other topics is very similar (~0.60). For perceived compliance, the means are more spread out (0.57 to 0.77). Also, the variances within the perceived compliance boxplots are higher than that within the actual compliance boxplots. Finally, the topic of system risk management has more outliers at the bottom end of compliance, for both actual and perceived compliance.

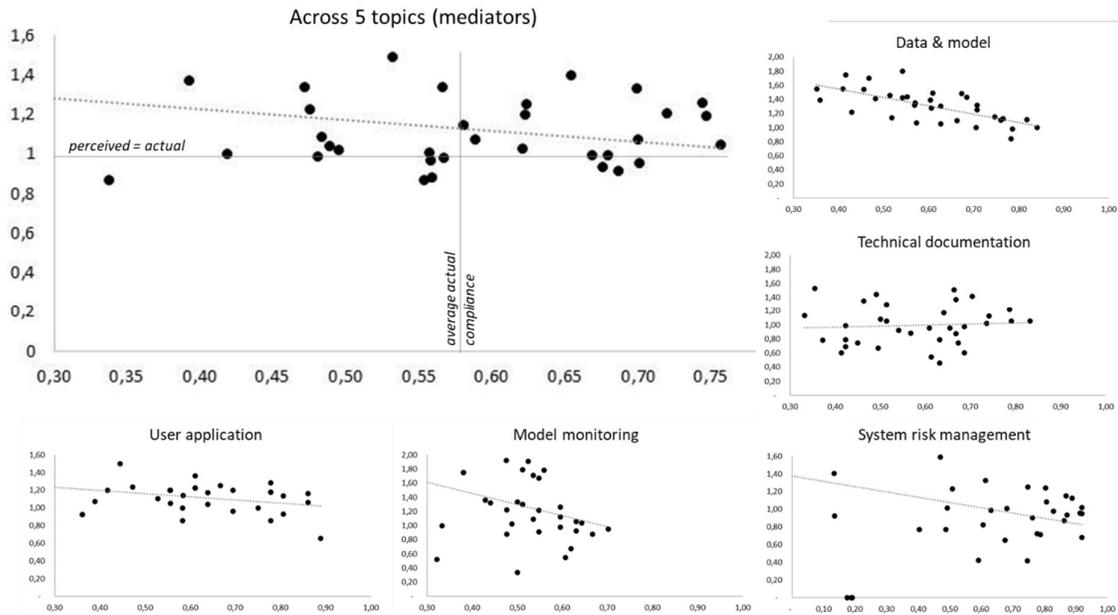

*Figure 4: Do participants exaggerate their compliance? (y-axes represent exaggeration ratio; x-axes represent actual compliance)*

### 6.3.2 Organizations exaggerating compliance

Figure 4 depicts exaggeration ratio versus actual compliance across and within the five topics. Almost all participants (~80%) exaggerated their compliance and except for technical documentation, the exaggeration ratio grows with declining actual compliance. Together with figure 3, it is also observed that in topics such as data & model and model monitoring exaggeration ratio on an average is greater than 1, whereas in topics such as technical documentation and system risk management it is less than 1. Together with a low actual compliance level, this could result in complacency, when complying with the AIA.

### 6.3.3 Effect of moderators

The ratio of participants demonstrating actual compliance per moderator is depicted in figure 5. The sample mean of actual compliance of 0,57 (sample median = 0,56) was chosen as a decision boundary for classifying compliance/non-compliance. Size of organizations as a moderator seems to influence compliance as is the case with familiarity with the AIA. The effect of insourcing/ outsourcing and in-shoring/offshoring is less evident.

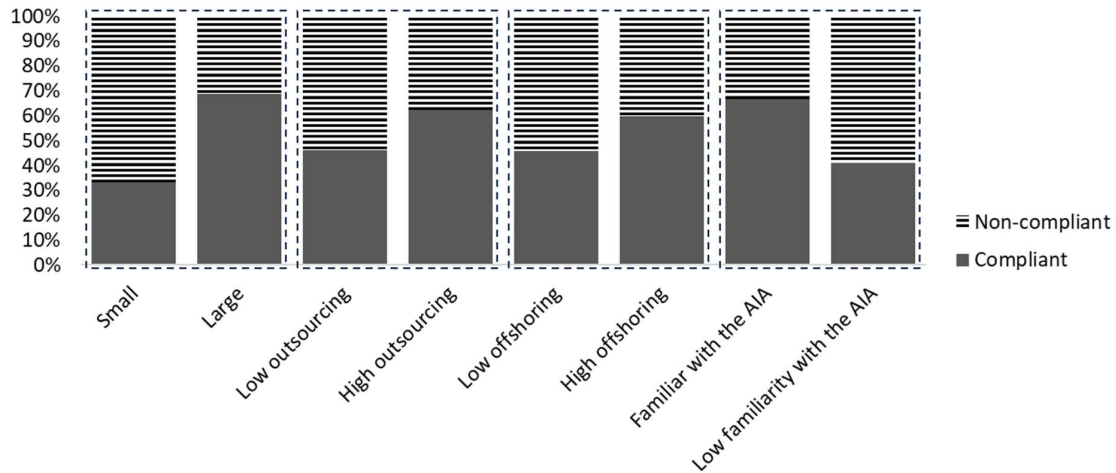

*Figure 5: Percentage of participants compliant per moderator*

To assess the significance statistically, hypotheses were formulated with the aim of performing chi-square tests of independence. The results are presented in table 3 below. For 3(a), $\chi^2$ = 3.841 with 1 degree of freedom. At a significance level of 95%, we reject $H_0$ implying that size and actual compliance are not independent of each other; smaller organizations demonstrate lower compliance. For the other tables (3b, 3c, 3d), it was not possible to conduct a test, as the minimum number of elements was not always 5.

*Table 3: Hypotheses and Chi-square tests of independence for moderators.*

| $H_0$: Size is independent of compliance | | | |
|---|---|---|---|
| $H_A$: Size is NOT independent of compliance | | | |
| | Compliant | Non-compliant | Total |
| Small | 6 | 12 | 18 |
| Large | 11 | 5 | 16 |
| Total | 17 | 17 | 34 |

(a)

| $H_0$: Outsourcing is independent of compliance | | | |
|---|---|---|---|
| $H_A$: Outsourcing is NOT independent of compliance | | | |
| | Compliant | Non-compliant | Total |
| Low outsourcing | 12 | 14 | 26 |
| High outsourcing | 5 | 3 | 8 |
| Total | 17 | 17 | 34 |

(b)

| $H_0$: Offshoring is independent of compliance | | | |
|---|---|---|---|
| $H_A$: Offshoring is NOT independent of compliance | | | |
| | Compliant | Non-compliant | Total |
| Low offshoring | 11 | 13 | 24 |
| High offshoring | 6 | 4 | 10 |
| Total | 17 | 17 | 34 |

(c)

| $H_0$: Familiarity with the AIA is independent of compliance | | | |
|---|---|---|---|
| $H_A$: Familiarity with the AIA is NOT independent of compliance | | | |
| | Compliant | Non-compliant | Total |
| High familiarity with the AIA | 8 | 4 | 12 |
| Low familiarity with the AIA | 9 | 13 | 22 |
| Total | 17 | 17 | 34 |

(d)

### 6.3.4 Regulatory pressure

As discussed in section 4, regulations such as the GDPR and the AIA have significant impact on organizations, often to varying extents depending upon their geography and size. A low priority to comply is also reflected in low actual compliance and vice versa. In figure 5 and table 3, it is concluded that size of an organization has a moderating effect on its actual compliance. Additionally, in figure 6, we notice that small organizations tend to have a lower priority to comply. It seems as if larger organizations endure more regulatory pressure and therefore, they have a higher internal priority to comply with the AIA.

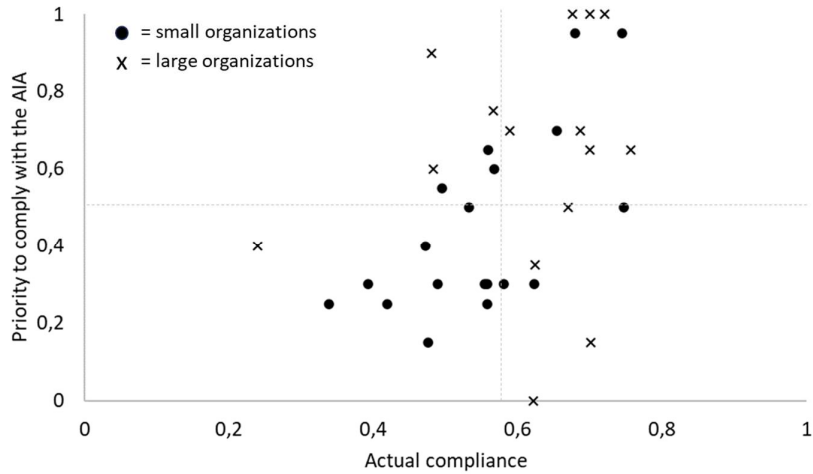

Figure 6: Regulatory pressure

A chi-square test of independence was subsequently conducted to assess the statistical significance of the relationship between organizations' internal priority to comply with the AIA and their actual compliance. The formulated hypothesis and the calculations are presented in table 4 below. The test statistic $\chi^2$ = 2.982 with 1 degree of freedom. At a significance level of 90%, we reject $H_0$ implying that the priority to comply with the AIA and actual compliance are dependant at a significance level of 10%.

Table 4: Hypothesis and Chi-square test of independence for priority to comply.

| $H_0$: Priority to comply with the AIA is independent of actual compliance | | | |
|---|---|---|---|
| $H_A$: Priority to comply with the AIA is NOT independent of actual compliance | | | |
| | Compliant | Non-compliant | Total |
| HIGH Priority to comply | 12 | 7 | 19 |
| LOW Priority to comply | 5 | 10 | 15 |
| Total | 17 | 17 | 34 |

# 7 Validation

The results presented in section 6 were deliberated upon in detail in two separate sessions with respondents from two organizations – a global leader in capital goods and home appliances (referred to as "Firm 1" below) and a startup focussed on developing AI systems for educational institutions (referred to as "Firm 2 below"). The two organizations were selected because they were very different from the perspective of the moderators: Firm 1 is a global recognized brand and market leader in engineering, whereas Firm 2 is a small local startup developing innovative AI systems for the education industry only. Firm 1 outsources and insources AI model development and production activities at onshore and offshore locations, with an emphasis on outsourcing and offshoring. Firm 2 only insources from its onshore location. Their actual compliance scores in the five topics of the survey lay far out of each other. The sessions were conducted online, recorded, and transcribed. The purpose of both sessions was two-fold: to validate the respondents' previous responses on the survey questions and to furthermore generate insights on their organizations' (lack of) compliance. The subsequent paragraphs present results of the two sessions.

## 7.1 Firm 1

The actual compliance score of Firm 1 was 0.75 and the exaggeration ratio was 1.04, indicating that the organization was not overstating its compliance. Within the five topics, data & model, technical documentation, user application, model monitoring and system risk management, Firm 1's actual compliance score was .79 (.60), .78 (.59), .89 (.60), .34 (51) and .93 (.60); the numbers in brackets reflect the average actual compliance in each topic. These scores and Firm 1's ranking relative to the participants initiated the validation discussion with a conscious effort to limit the discussion to the organizational paradigm presented in figure 1 and to a specific product group only. Given the large size of Firm 1, the latter was necessary to keep the discussion focussed. The results of the validation discussion can be summarized as 3 types of challenges Firm 1 faced: requirements specification, resource-related and process-related.

### Requirements specification challenges

In general, within Firm 1 communication about the AIA is best described as informal and tacit. The team working on developing AI models is aware of its existence and is speculative about its potential implications on their activities. The speculation appeared to be enabled by every individual's best-effort approach towards AIA compliance driven by ethics and individual moral responsibility (Kormelink, 2019) and their concerns that when enforced by regulatory bodies, the AIA could have significant impact on their activities. Furthermore, in the absence of training, formal guidelines, and procedures, which is also reflected in the survey response, model developers believe that their skills and best-efforts could be inadequate in ensuring that functional requirements are incorporated in their AI systems in a manner that these systems are compliant with the AIA. The top-5 challenges faced could be summarized as:

- Ai. internal non-alignment over how AIA compliance requirements will affect functional and technical requirements at a product level,
- Aii. lack of consensus on interpreting AIA compliance requirements driven by individual ethical and moral responsibility perceptions,
- Aiii. risks from lack of AIA compliance and business value of AIA systems being too often pitted against each other with hardly any trade-offs,
- Aiv. compliance requirements being uncoordinated during handovers (e.g., across organizational boundaries), and,
- Av. concerns over inadequate competency to ensure that requirements are AIA compliant.

### Resource challenges

Despite being a recognized brand and a preferred employer, Firm 1's resource challenges are representative of the short tenure of data professionals (Eastwood, 2022). Even at its inshore location, majority of its data professionals are foreign nationals with an average tenure of less than 2 years. These professionals leave Firm 1 for jobs as data professionals elsewhere, driven by better employment terms and conditions. This poses three challenges. First, providing training under these circumstances is considered infeasible. Second, frequent handovers put strain on their quality and the continuity of projects. Lastly the different nationalities and cultural backgrounds within the same team, result in sub-optimal alignment on ethical issues and individual moral responsibilities. This was exacerbated by the absence of repositories aimed at managing and demonstrating compliance of AI-systems and the lack of a single point of contact on compliance & ethical issues related to AI systems. The top-5 challenges faced could be summarized as:

- Bi.     high churn rate among data professionals that makes training infeasible,
- Bii.    too frequent handovers straining project quality and continuity,
- Biii.   divergent values on ethics and individual moral responsibility,
- Biv.    lack of a single point of contact on AIA compliance issues and ethics of AI systems, and,
- Bv.     absence of a repository for managing and demonstrating compliance.

Process challenges

Being a global market leader in engineering, Firm 1 conducts its operations through well-documented processes that comply to relevant ISO standards. From the perspective of business process maturity, it is at least an optimized enterprise (Fisher, 2004). It is not surprising therefore that it scored high on technical documentation, user application, and system risk management. However, when it comes to processes that govern, describe and support model monitoring, Firm 1 does not quite resemble its own self, given its actual compliance score of 0.34. This had to do with two challenges that Firm 1 presently confronts. First, the drive to AI is driven primarily by acquisitions made by the organization. Integration of the acquisitions are often delayed (due to planned and deliberate reasons), which leads to postponed roll-out of Firm 1's organization-wide processes in the newly acquired target. The target operates siloed (Fisher, 2004). This impacts model-monitoring disproportionately since management attention post-merger lies on value-capture through synergies, thereby overshadowing the monitoring of legacy models for (continued) compliance. Second, and probably more important, Firm 1 has no established AI compliance processes in place. This is reflected in its below-average compliance of the risk management questions of the survey, indicating that risk mitigating actions of AI systems are not put in the context of organization-wide compliance processes. Much awaits legislative developments around the AIA. Looking forward, data professionals are curious how this will affect their daily work and how compliance processes in the future affect their primary process of product development. To what extent this eventually distracts data professionals from their current daily activities, lead to increased overhead and compromise the exploratory nature of their AI development activities are also concerns that emerged during the validation discussion. The challenges can be summarized as below:

- Ci.     non-integration of acquired target organizations making them operate more as stand-alone units,
- Cii.    undefined and/or informal processes related to AI compliance,
- Ciii.   individual best-efforts often leading to disrupted workflows and drop in productivity,
- Civ.    concerns among data professionals that new AI compliance processes will result in additional steps in their workflows and therefore increase overhead, and,
- Cv.     concerns that innovation culture and mindset of the team will suffer from too much of compliance-control.

It was expected that given its global reputation, Firm 1 would respond to regulatory pressure (Malesky & Taussig, 2017) by initiating process development initiatives to comply to the upcoming AIA. Instead, by deciding to secure the value of an acquisition through non-integration, it succumbed to financial pressure (von Zahn et al., 2021). On a positive note, the concerns on non-compliance among Firm 1's data professionals are driven by normative pressures (Durand et al., 2019) from compliance developments in other parts of Firm 1 and adjacent legislations such as GDPR. Another positive aspect of the situation was the explorative nature of innovation, which data professionals highly enjoyed at Firm 1.

## 7.2 Firm 2

Firm 2's actual compliance score was 0.57 and the exaggeration was at 1.34, indicating that the organization was highly overstating its compliance. Its actual compliance score was .46, .59, .61, .51 and .70 in data & model, technical documentation, user application, model monitoring and system risk management. As with Firm 1, these observations were used to initiate the validation discussion.

Since Firm 2 is a startup with a very modest organization, its employees are expected to be multifunctional, be willing to learn new skills, and deal with a broader range of activities. Firm 2 is steered more by technology and financial pressures than by regulatory ones. Bluntly put, meeting budgetary targets, and delivering on product promises preoccupies management agenda as opposed to AIA compliance. 'Avoidance' and 'compromise' as partial compliance strategies (Oliver, 1991) and possibly 'creative compliance' are observed during the validation discussion. This has resulted in data professionals embracing the task of incorporating AI compliance in Firm 2's products without any well-defined processes and dedicated resources.

The single-mindedness of demonstrating innovative products results in basic activities of building a compliant AI-system being neglected, thereby the substantiating the low actual compliance score in data & model. The financial and technology pressures often result in developing systems partially in academic contexts, on which Firm 2 has very limited influence to ultimately ensure compliance. Professionals have neither the competencies to test model/data bias nor the time to train themselves and acquire these competencies. This could result in AI-systems brought to the market without prior knowledge and consideration of the harmful effects of the system.

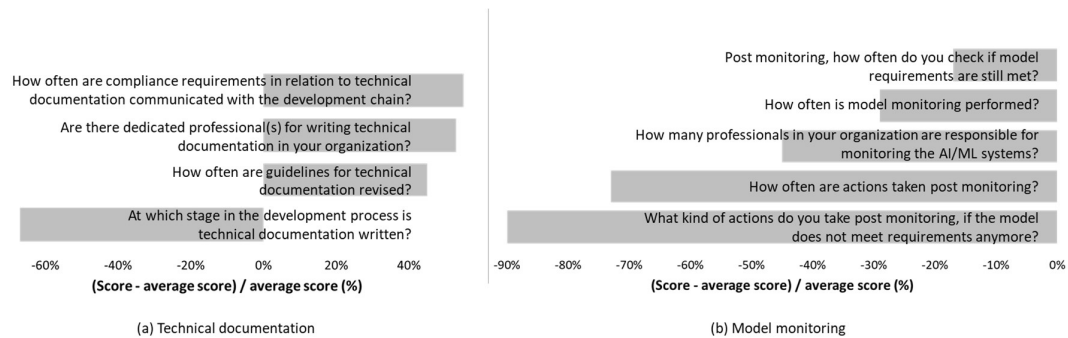

*Figure 7: Responses to questions within technical documentation (a) and model monitoring (b) by Firm 2.*

The "compromise" attitude towards compliance is noticeable also in figure 7a. Firm 2's compliance score in technical documentation matches the average (0.59); almost all the technical documentation activities are performed in a way that is above the average score of the survey respondents; however, these activities are not executed frequently enough to assure and demonstrate compliance (Baldwin & Cave, 1999). Firm 2's responses within model monitoring (figure 7b) reaffirm the rationale of "compromise": the response on the questions becomes more negative as the questions become more specific. This is also reflected in Firm 2's high exaggeration ratio of 1.34. Building upon the challenges mentioned in section 7.1.1, Firm 2's challenges can be summarized as below:

    Avi.    blind spot towards compliance requirements due to competing technology and financial pressures,

Bvi.     lack of dedicated compliance specialists resulting in insufficient comprehension of the AIA and its inclusion in the organization's way of working, and,

Cvi.     undefined processes or processes defined at too high levels, which make defining and introducing new AIA compliance processes confronting.

# 8   The APPRAISE framework

The empirical research and validation interviews present insights on how organizations, confronted with the four pressures, struggle to achieve AIA compliance. The following sections describe the complexity at a strategic level and present rationale for the proposed APPRAISE governance framework for AI innovation (figure 8).

## 8.1   The governance complexity

Compliance in the organizational paradigm in figure 1 presents several challenges. First, the scope is broad and interdisciplinary. AI developers, legal experts and commercial professionals have limited knowledge in each other's domains (Schultz, 2013). Second, the number of organization interfaces that need to be coordinated is high, especially when combined with outsourcing and offshoring. Third, coordinating control objectives and business objectives within similar interdisciplinary processes is often complex (Sadiq et al., 2007). Fourth, rigid compliance processes may come at a cost of innovation, especially in exploration contexts (section 3), and hence, impact value creation (section 7). Lastly and probably the most complex challenge is that the four pressures are omnipresent in strategic value creation. This is best visualized using a strategy map (Kaplan & Norton, 2000).

Findings from secondary research presented in sections 2, 3, 4, and 5 elucidate this last challenge on the strategy map. Value creation pressure has a first order impact on financial perspective through improved operational efficiencies and enhanced customer value that AI systems typically generate, such as with recommender systems. The exploration/exploitation and build/buy dilemmas have impact, among others, on processes in the internal perspective and on leadership in the learning & growth perspective. Technology pressure impacts customer perspective, for example, through additional and improved functions in existing and new products/services. It also impacts the internal and learning & growth perspectives owing to innovation speed being critical in processes, culture, and teamwork. Similarly, the effect of regulatory pressure is visible in regulatory processes, leadership, and alignment. Normative pressure impacts learning & growth perspective most, however, its impact is also felt in how innovation and coordination processes are designed. These challenges and resulting complexities that organizations encounter at strategic and tactical levels aggregates to one fundamental question: how should AI-driven innovation be governed?

## 8.2   Importance of auditability

Success of legislation and regulation efforts rely heavily on enforcement mechanisms, which differ based upon, e.g., the selected regulation strategy (Baldwin & Cave, 1999). In a 'Command and Control' regulatory strategy, highly reflected in the AIA approach, a good balance between deterrence and compliance is essential to avoid 'creative compliance' and stimulate good dialogue and information exchange between notified bodies and providers of high-risk stand-alone AI systems. The importance of audit has also been emphasized also from an ethical perspective (Rai, 2022). Auditing the ethics of AI systems encompasses deontological and consequential ethics (Bringas Colmenarejo et al., 2022), one in which the consequential nature and process of developing an AI system needs to be audited along with the decision-making process at a provider (Finocchiaro,

2023). This will not only ensure that the causal developmental process is robust, but also stimulate correctness, trustworthiness, and the right culture at providers.

While ensuring quality of AI systems, the purpose of such an audit is three-fold: first, to raise the probability of both discovering and reporting a breach (DeAngelo, 1981); second, a structured mechanism on the basis of which the past and present behavior of an organization can be offset against a set of requirements (Mökander et al., 2022), and, third, achieve a sustained homogenous impact, which is yet to be observed under the GDPR. Consequently, enforcement through audit would result in demonstrable compliance (Hoepman, 2014) and good governance.

## 8.3   The modules and their rationale

The APPRAISE framework connects strategic variables identified in this paper to the proposed modules, coordination, organization capital, technology safety, and provider alignment thereby aiming to create responsible value and enable auditability. Each of the four comprises of a non-exhaustive list of attributes. Provider alignment and technology safety have a lower order impact on AIA compliance and responsible value creation as compared to coordination and organization capital. Subsequent subsections present rationale behind these building blocks. The importance of organizational size, build/buy, and exploitation/exploration dilemmas as strategic variables has been presented earlier in sections 2, 3, and 6. The impact of strategic decisions taken in these variables (and possible more) is felt in the modules.

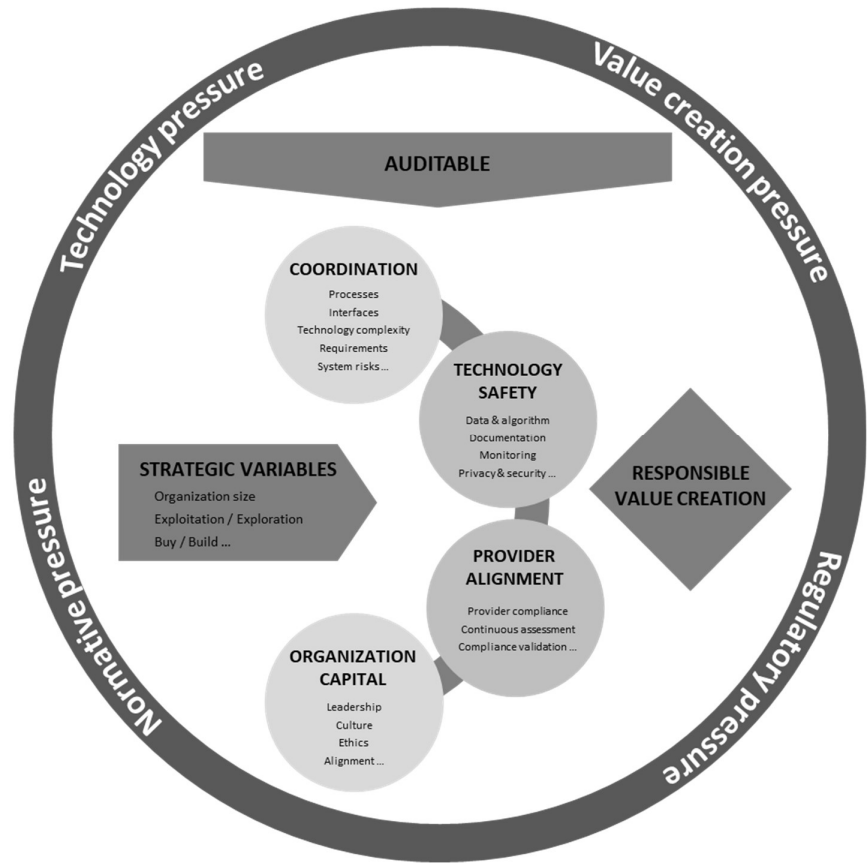

*Figure 8: The APPRAISE governance framework for AI innovation*

### Coordination

Multiple needs underscore the importance of coordination: first, a dual approach towards business and compliance processes (section 8.1), second, varying process-rigidity (section 3); third, a potentially large number of organizational interfaces resulting from build/buy decisions (section 3); fourth, technology complexity aggravated by the need for speed (sections 2, 6.3); fifth, a dual approach towards functional/technical requirements (sections 6.1, 7.1); and, lastly, system risks that surface from every nook and corner. For audit, KPIs for coordination is a topic of future research. However, parallels could be drawn from the global software engineering industry (Bhat et al., 2006), in which key reported issues process variations (Espinosa et al., 2007), variations in development practices & styles (Smite et al., 2008), communication gaps (Bhat et al., 2006) and lack of trust (Moe & Smite, 2008) across boundaries (Nguyen-Duc & Cruzes, 2013).

### Organization capital

The importance of organization capital has been presented earlier through its attributes, such as leadership, culture, and ethics in sections 5 and 7. Alignment, essential to churn out faster and better innovations, is presented in section 2. Despite the pessimistic public tone towards auditing culture (Furnham & Gunter, 1993), research on how these elements of organization capital impact creativity & innovation and performance (Naranjo-Valencia et al., 2016) are abundant and increasingly common practice (Tian et al., 2018). Additionally, as AI systems behave increasingly as black boxes, the role of IMR becomes more significant. Using these elements and the strategic variables as a starting point, organizations must create assessment frameworks to audit their organization capital. The objectives of audit (section 8.2.1), when fulfilled will in turn deliver responsible value creation.

### Technology safety

The attributes of technology safety are a direct outcome of the AIA (table 1) and discussed in length in section 6. Technically assessing AI systems for explainability is great research interest (Miller, 2019) today. Examples include methods for auditing binary classification algorithms (Bhaumik & Dey, 2023) or recommender systems (Chen et al., 2023). Such studies propose technical audit frameworks aimed at multiple aspects including, discrimination and fairness. Auditing technical requirements for privacy and security is abundant in literature (Awaysheh et al., 2021).

### Provider alignment

Multiple software systems working together in a cloud-based environment is today common practice. This is the outcome of the build/buy and the exploration/exploitation dilemmas discussed earlier. The AIA requires providers of such systems to be compliant and necessitates downstream organizations deploying such systems to demonstrate compliance (section 4). Therefore, the latter must oblige the former to demonstrate compliance of its AI systems. Additionally, the value-chain of a specific provider could comprise of multiple sub-providers, who may perform updates at regular intervals or driven by specific events. In this dynamic environment, managing risks continuously necessitates agreements with providers that facilitate an acceptable level of risk on behalf of the focal organization. Depending upon the level of impact a provider's system has on a user and the provider's risk profile, the focal organization should request validation of the provider's AI system compliance. The focal organization needs to formulate assessment frameworks aimed at managing and evaluating its relationship with its providers.

Although attempts to connect corporate social responsibility to value has often been described as a scrimmage (Nguyen et al., 2020), the extent to which organizations embrace these modules and audit them will greatly determine their level of compliance and responsible value creation. The importance of ethical management of AI has been researched by other frameworks, e.g., within the EMMA framework (Brendel et al., 2021), in which managing AI ethics has been conceptualized as an interplay of managerial decisions, ethical considerations and environmental dimensions. Combining these modules to strategic influence of value in corporate decision-making, the proposed framework enables responsible value creation.

## 9   Discussion & conclusions

To start with, this paper presents four pressures that organizations currently experience when embracing AI for product/service innovation. These include technology, value creation, regulatory and normative pressures. Organizations react differently to these pressures depending upon their size, their familiarity with regulation (AIA), and the choices they make in their build/buy and exploration/exploitation dilemmas. The latter manifests itself in the level of outsourcing and offshoring that an organization undertakes to fulfil its AI ambitions. Such strategic choices have an impact on organizations' ability to create value responsibly and eventually on their level of compliance with the AIA.

Through an empirical study involving 34 organizations in the Netherlands, valuable insights are obtained on the AIA compliance that organizations demonstrate. The average compliance score was low and almost all participants overstated their compliance in the survey. Smaller organizations demonstrated lower compliance. Also, organizations with a low familiarity with the AIA tend to be less compliant. The effect of offshoring and outsourcing, although indicative, did not provide any conclusive results. An organization's actual compliance is also dependent on the internal priority that it places on complying with the AIA and smaller companies tend to place lesser priority in complying with the AIA. The heterogenous moderating effect of organization size on GDPR's success is also noticeable in the case of the AIA.

The validation meetings provided deeper insights into root causes. They depict organizations performing balancing acts among the four pressures. They reveal the resource and process challenges organizations encounter and the importance of co-ordination. Creative compliance and partial compliance was also observed. Subsequently, the governance complexity has been established and the case for audit presented. The AIA, being more complex and ambitious than the GDPR, should use audit as an enforcement mechanism. Based on primary and secondary research, this paper proposes the APPRAISE governance framework for AI innovation bridging the gap between strategic variables and responsible value creation, with auditability as a constraint.

In developing the framework, the focus was capturing the breadth and structuring the governance challenge that organizations face in creating responsible value with AI. There are many areas that require further research to improve the reliability of the insights in this paper and generate additional insights. For example, the moderating effect of outsourcing and offshoring needs further investigation as much as the effect of combining moderators. This would add to the body of knowledge of how moderators play roles in AIA compliance. Furthermore, the modules presented in the framework need research, especially with relation to the moderators and the four pressures. This is expected to result in detailed assessment frameworks, which organizations could eventually use to audit themselves.